\documentclass[slac_one]{revtex4}
\usepackage{graphicx}
\usepackage{fancyhdr}
\begin{document}

\title{Inclusive Diffraction at HERA} %% Paper title goes here


\author{Paul Laycock (for the H1 and Zeus collaborations)}
\affiliation{University of Liverpool - Dept of High Energy Physics \\
Oliver Lodge Laboratory, Liverpool L69 7ZE, UK}

\begin{abstract}

  The H1 and Zeus collaborations have measured the inclusive
  diffractive DIS cross section $ep \rightarrow eXp$ and these
  measurements are in good agreement within a normalisation
  uncertainty.  Diffractive parton density functions (DPDFs) have been
  extracted from NLO QCD fits to inclusive measurements of diffractive
  DIS and the predictions of these DPDFs are compared with
  measurements of diffractive dijets in DIS, testing the validity of
  the factorisation approximations used in their extraction.  H1 then
  use these diffractive dijets in DIS data to provide further
  constraints in a combined QCD fit, resulting in the next generation
  of DPDFs which have constrained the diffractive gluon at large
  momentum fractions.  Finally, the predictions of DPDFs are compared
  to diffractive dijets in photoproduction where the issue of survival
  probability in a hadron-hadron environment can be studied.
\end{abstract}

\maketitle

\thispagestyle{fancy}

\section{Diffraction at HERA}

\begin{figure}[h]
\begin{center}
\includegraphics[width=0.4\columnwidth]{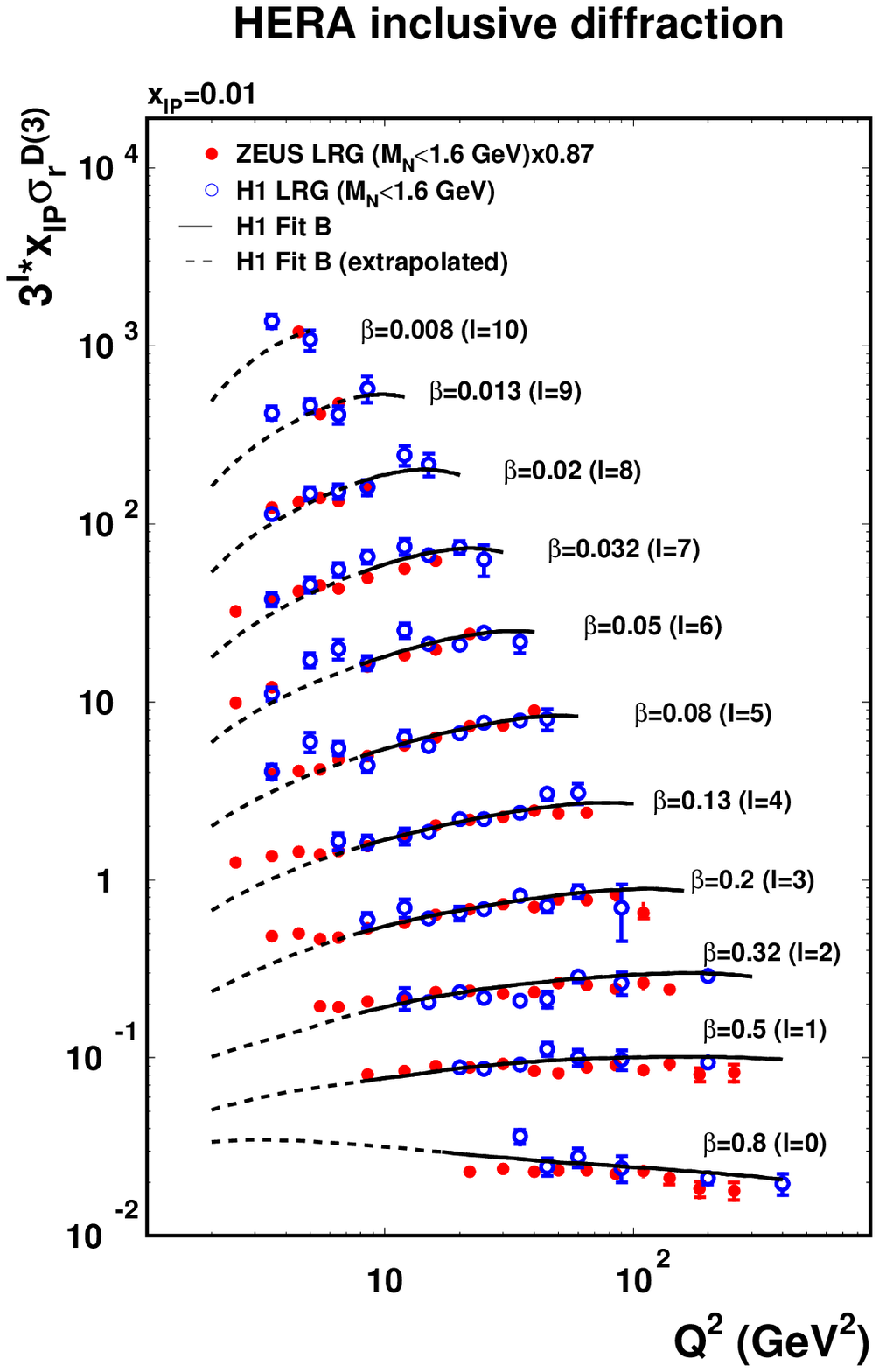}
\includegraphics[width=0.4\columnwidth]{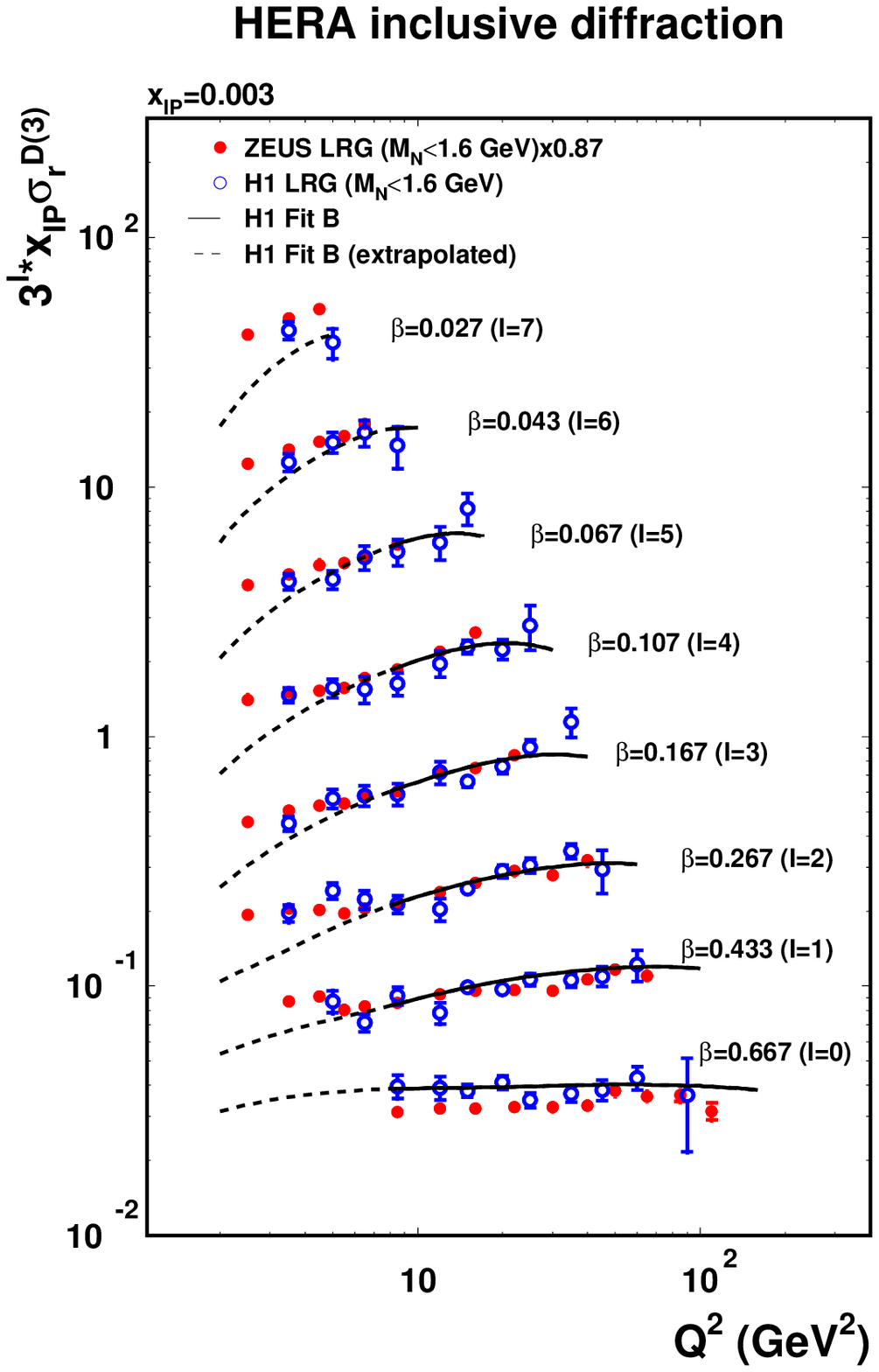}
\caption{The reduced diffractive cross section as measured by the H1 and Zeus collaborations.}
\label{Fig:sigma}
\end{center}
\end{figure}

It has been shown by Collins~\cite{Collins} that the NC diffractive
DIS process $ep\rightarrow eXp$ at HERA factorises; a useful
additional assumption is often made whereby the proton vertex dynamics
factorise from the vertex of the hard scatter - proton vertex
factorisation.
The kinematic variables used to describe inclusive DIS are 
the virtuality of the exchanged boson $Q^2$, the
Bjorken scaling variable $x$ and $y$ the inelasticity.
In addition, the kinematic
variables $x_{I\!P}$ and $\beta$ are useful in describing the
diffractive DIS interaction.  
$x_{I\!P}$ is the longitudinal fractional momentum of the proton
carried by the diffractive exchange and $\beta$ is the longitudinal
momentum fraction of the struck parton with respect to the diffractive
exchange; $x=x_{I\!P}\beta$.  The data are discussed in terms of a
reduced diffractive cross-section, $\sigma_r^{D(3)}(\beta, Q^2,
x_{I\!P})$, which is related to the measured differential cross
section by:
\begin{equation}
\frac{d^3\sigma_{ep \rightarrow eXp}}{d\beta dQ^2 dx_{I\!P}} = \frac{4\pi\alpha_{em}^2}{\beta Q^4}(1 - y + \frac{y^2}{2})\sigma_r^{D(3)}(\beta, Q^2, x_{I\!P}).
\end{equation}
In the proton vertex factorisation scheme, the $Q^2$ and $\beta$
dependences of the reduced cross section factorise from the $x_{I\!P}$
dependence.  Measurements of the reduced diffractive cross section
from both H1 and Zeus are shown in Figure~$\ref{Fig:sigma}$, where the
new Zeus preliminary measurement has been scaled by a factor of 0.87,
a factor consistent with the normalisation uncertainties of the two
analyses.  The measurements agree rather well.

\subsection{Diffractive PDFs from Inclusive data}

Using the approximation of proton vertex factorisation, the H1 and
Zeus collaborations have extracted DPDFs using NLO QCD fits to the
$\beta$ and $Q^2$ dependencies of the reduced cross
section~\cite{H1Inc, Zeus}.  H1 obtained two fits of approximately
equal quality, Fit A and Fit B, differing only in the number of terms
used to parameterise the gluon.  The two fits, while fully consistent
at low fractional momentum, yield very different results for the
diffractive gluon at high fractional momentum.  This is due to
quark-driven evolution dominating the logarithmic $Q^2$ derivative of
the reduced cross section at high $\beta$, which in turn greatly
reduces the sensitivity of this quantity to the gluon.

\section{Diffractive dijets in DIS}

\begin{figure}[h]
\begin{center}
\includegraphics[width=0.5\columnwidth]{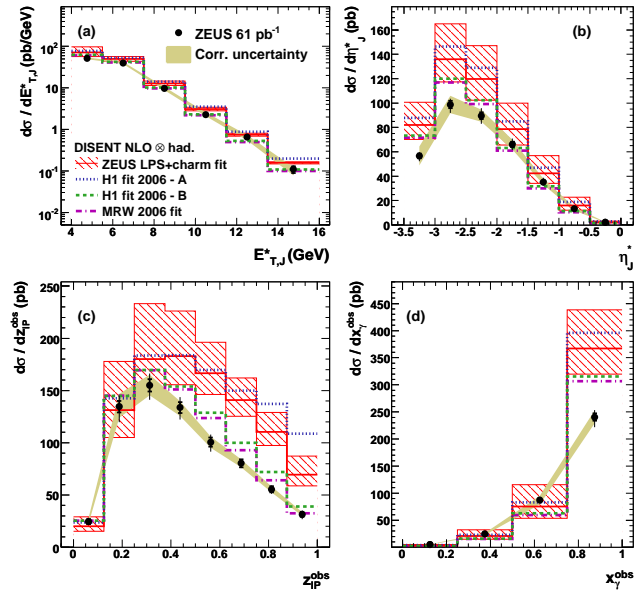}
\caption{The diffractive dijets in DIS data compared to the predictions of fits to inclusive DIS data.}
\label{Fig:comp}
\end{center}
\end{figure}

Diffractive dijets in DIS provide a sensitive experimental probe of
the diffractive gluon, as the dominant production mechanism is
boson-gluon fusion.  The sensitive variable is $z_{I\!P}=\frac{Q^2 +
  M_{12}^2}{Q^2+M_X^2}$, where $M_{12}$ is the invariant mass of the
dijet system.  Both H1 and Zeus have measured the diffractive dijet
cross section in DIS~\cite{Dijets, Zeus_Dijets}.  In
Figure~$\ref{Fig:comp}$, the Zeus measurement is compared to the
predictions of a Zeus fit to inclusive data and H1 Fit A and Fit B.
At low $z_{I\!P}$, where the inclusive data have sensitivity to the
diffractive gluon, the results of the predictions are very similar and
agree well with the data.  This supports the use of the proton vertex
factorisation approximation needed to make the NLO QCD fits.  At high
$z_{I\!P}$ the data clearly prefer the prediction of Fit B.

\begin{figure}[h]
\begin{center}
\hspace{-1.0cm}
\includegraphics[width=0.15\textwidth]{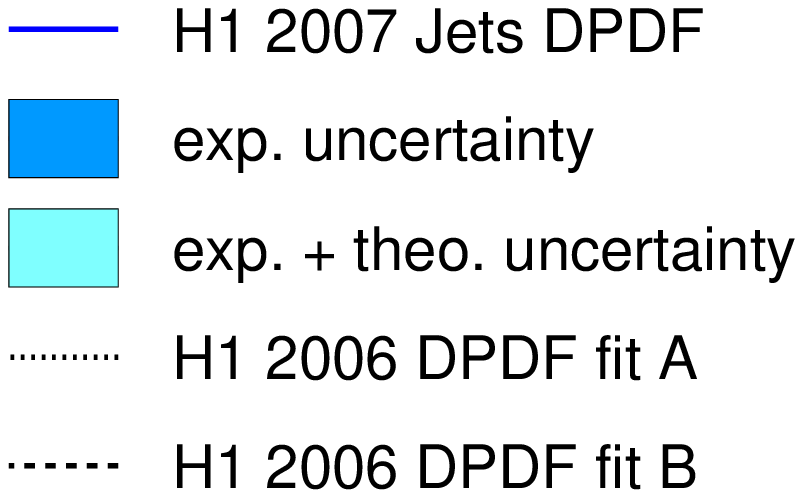}
\hspace{12.0cm}
\vspace{0.1cm}
\includegraphics[width=0.25\textwidth]{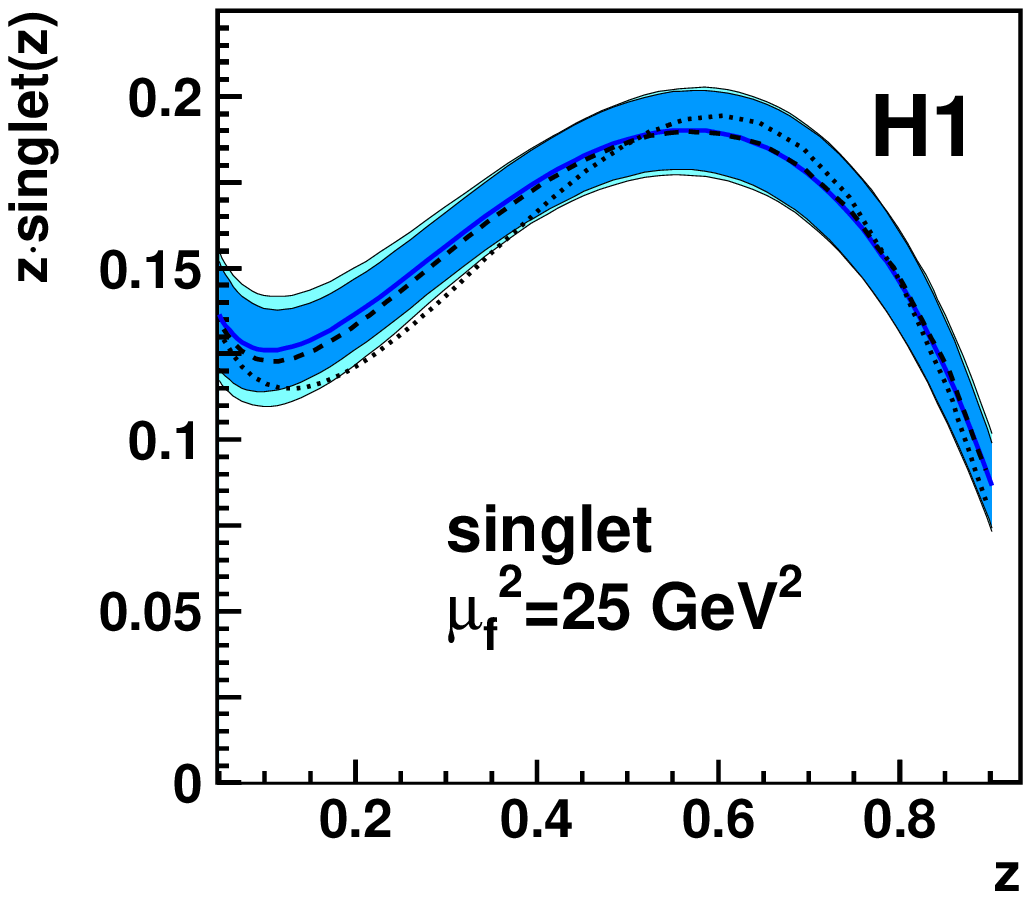}
\includegraphics[width=0.25\textwidth]{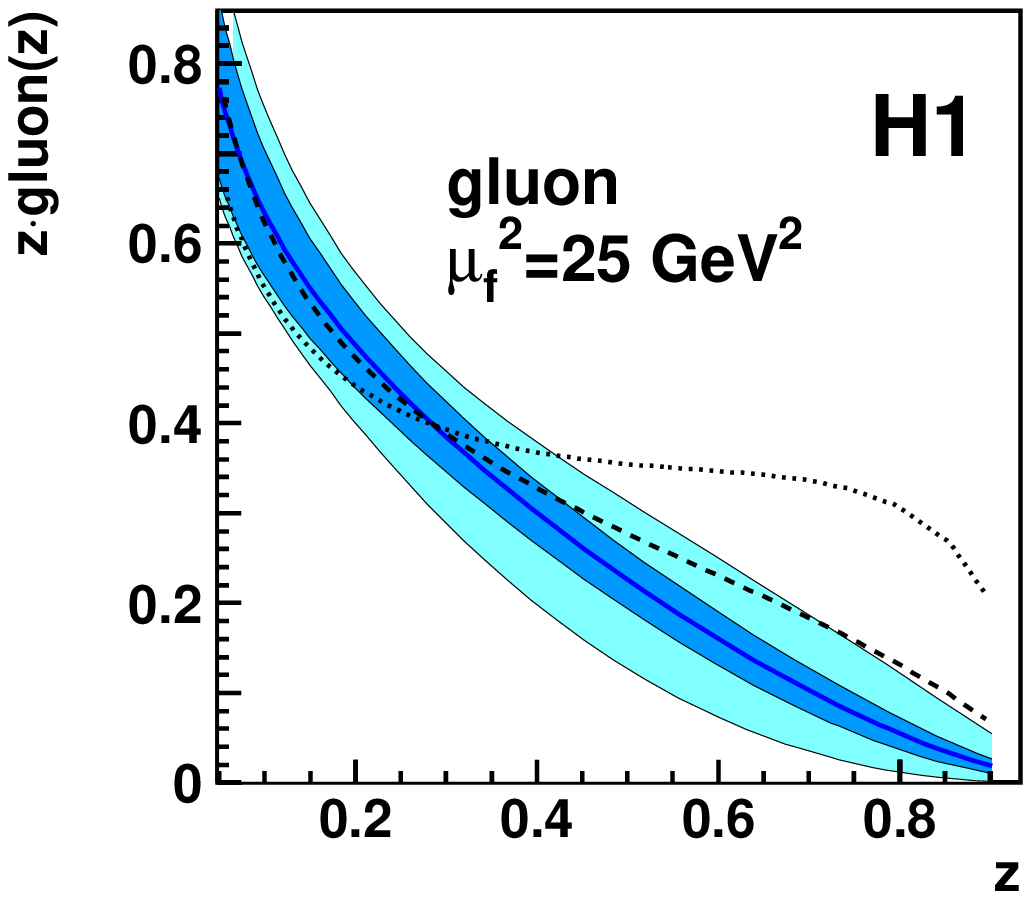}
\hspace{5.0cm}
\vspace{0.1cm}
\includegraphics[width=0.25\textwidth]{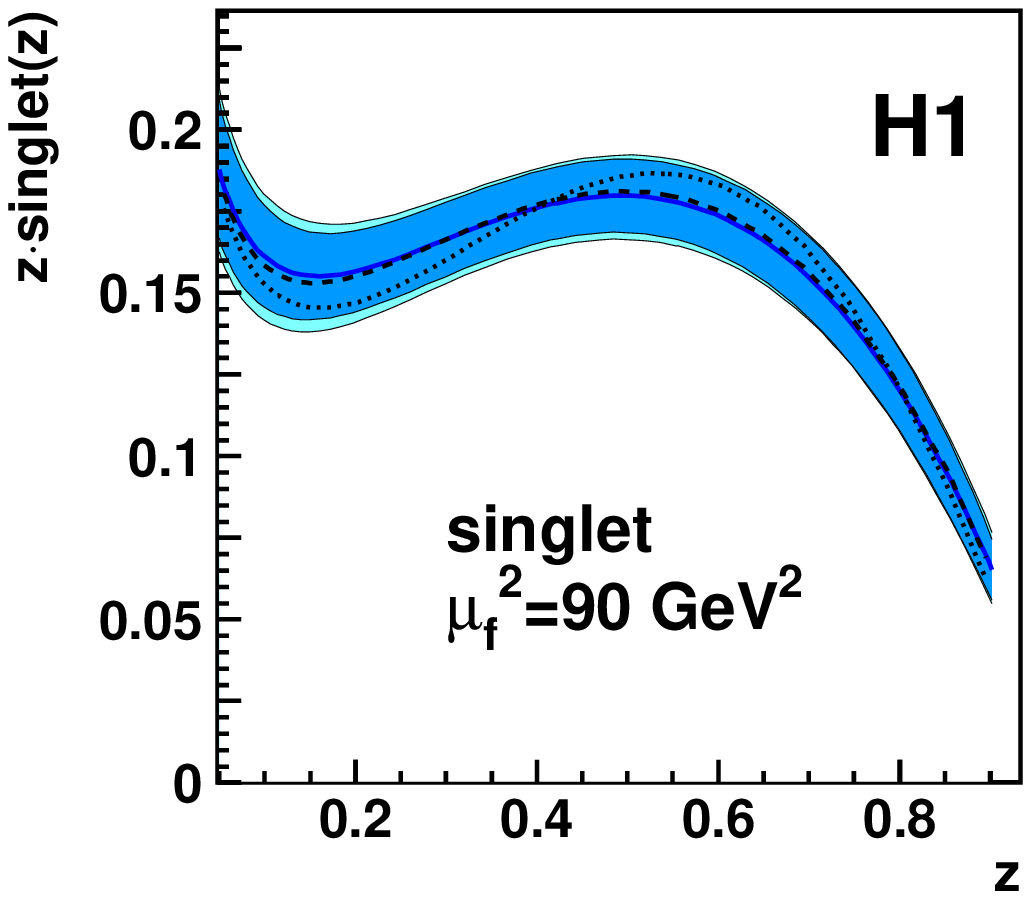}
\includegraphics[width=0.25\textwidth]{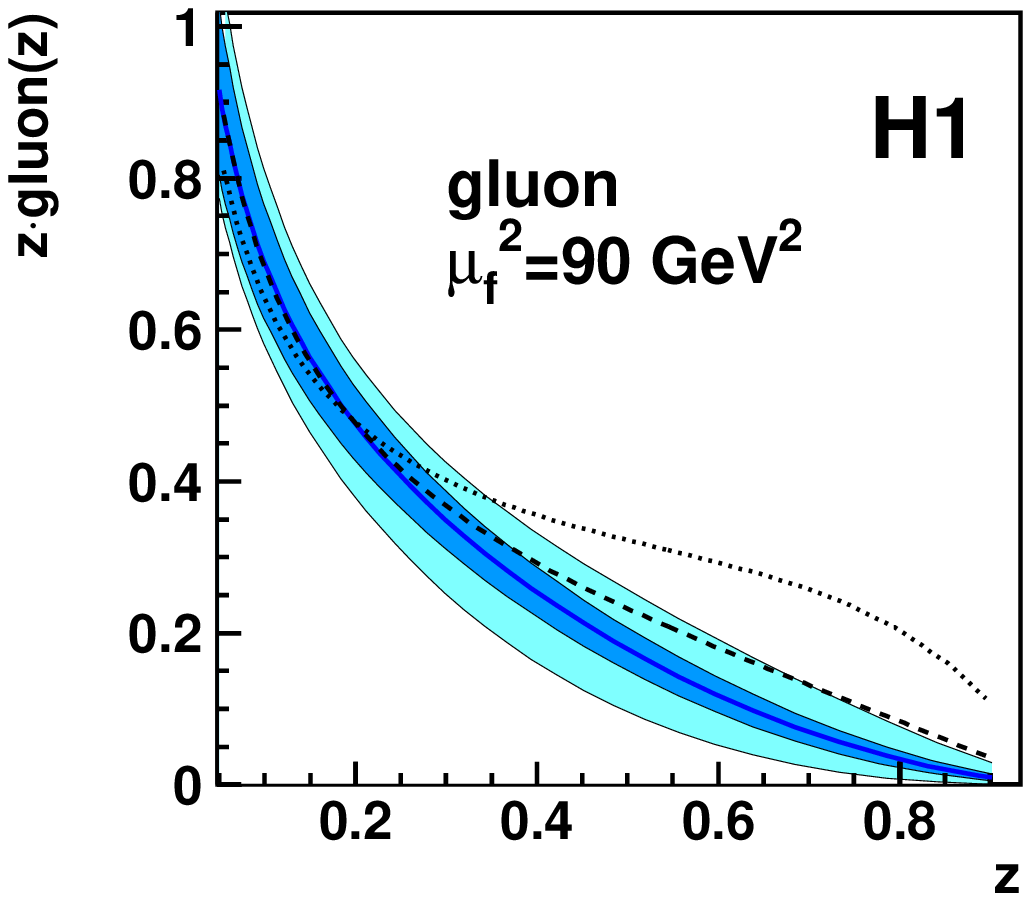}
\end{center}
\caption{The H1 DPDFs resulting from 
the combined fit to the inclusive and dijet diffractive DIS data.}
\label{Fig:JetDPDFs}
\end{figure}

Having shown the sensitivity of the diffractive dijets in DIS data, H1
have included their data in a combined fit with the inclusive
diffractive DIS data~\cite{Dijets}.  The resulting fit is
indistinguishable from Fit A and Fit B in its description of the
inclusive data and produces a better description of the diffractive
dijet data, consistent with that of Fit B.  The resulting DPDFs from
this combined fit, are shown in Figure~$\ref{Fig:JetDPDFs}$.  Both
singlet and gluon are constrained with similar good precision across
the whole kinematic range.

\section{Diffractive dijets in photoproduction}

\begin{figure}[h]
\begin{center}
\includegraphics[width=0.28\columnwidth]{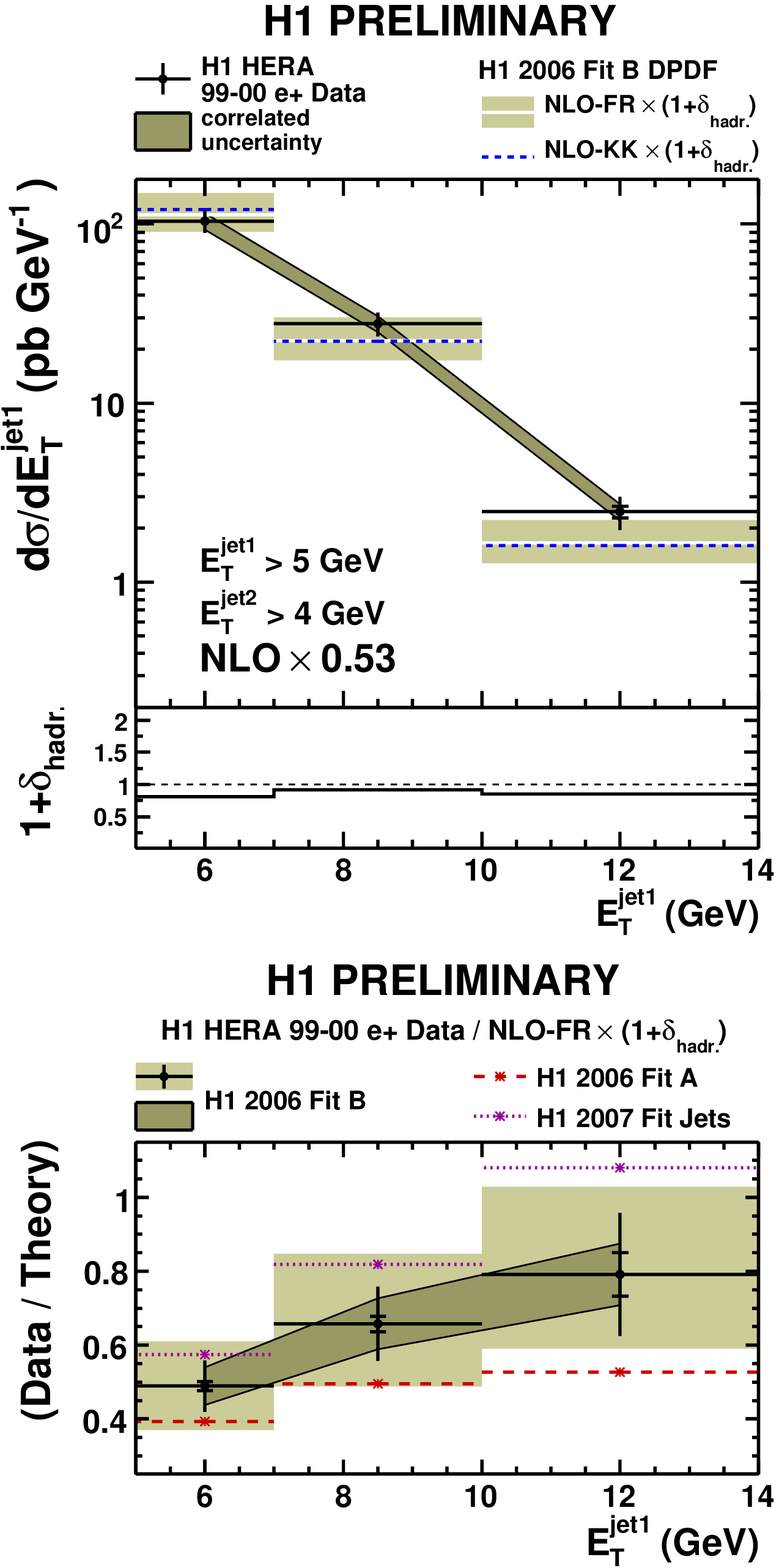}
\includegraphics[width=0.28\columnwidth]{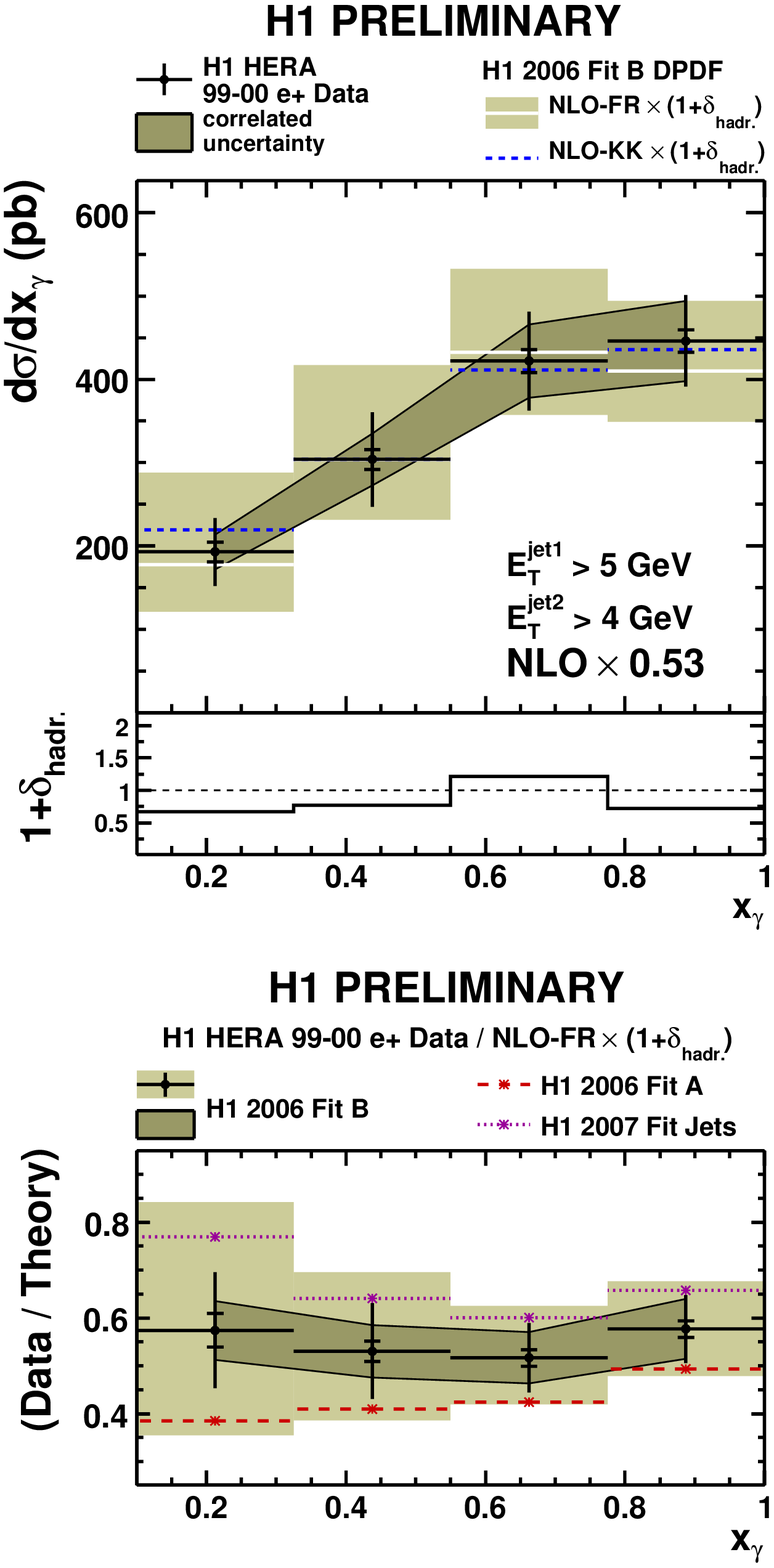}
\caption{The diffractive dijets in photoproduction data compared to the predictions of fits to inclusive DIS data and a combined fit to inclusive and dijet data.}
\label{Fig:ppcomp}
\end{center}
\end{figure}

Despite the success of factorisation in diffractive DIS at the HERA
experiments, there is a long-standing issue that the predictions
obtained with HERA DPDFs grossly overshoot the diffractive dijet cross
section at the Tevatron.  At HERA, photoproduction events, where
$Q^2\sim0$, provides an environment similar to a hadron-hadron
collider.  The variable $x_{\gamma}$ is the fraction of the four
momentum of the photon transferred to the hard interaction; the lower
the value of $x_{\gamma}$ the more hadron-like the photon.  Both H1
and Zeus have measured diffractive dijets in
photoproduction~\cite{Sebastian, Zeus_dijetsingammap}. The latest
preliminary results from H1~\cite{Karel} are shown in
Figure~$\ref{Fig:ppcomp}$ compared to the predictions of Fit A and Fit
B and the prediction of the combined fit including dijet data
described above.  There is a suppression of the cross section with
respect to the predictions and this suppression is independent of
$x_{\gamma}$.  There is also a suggestion that this suppression is
dependent of the $E_T$ of the jet.  This would be consistent with the
Zeus analysis at higher $E_T$ where less suppression is observed.

\section{Conclusions}

The H1 and Zeus collaborations have measured the inclusive diffractive
DIS cross section $ep \rightarrow eXp$ and these measurements are in
good agreement within their normalisation uncertainties.  The DPDFs
from NLO QCD fits to inclusive diffractive DIS data can predict the
diffractive dijets in DIS cross section at low $z_{I\!P}$ while at
high $z_{I\!P}$ the data favour Fit B.  Including the diffractive
dijet data in a combined fit further constrains the NLO QCD fit, where
the inclusive data alone are unable to unambiguously constrain the
diffractive gluon.  The resulting DPDFs from H1 are constrained with
good precision across the whole kinematic range.  Finally, when
compared to the predictions of DPDFs, diffractive dijets in
photoproduction show a suppression of the cross section which is
independent of $x_{\gamma}$ but which is consistent with an $E_T$
dependence.

\bibliographystyle{unsrt}
\bibliography{ichep08_laycock}

\end{document}